\documentclass[prl,aps,epsf]{revtex4}
\usepackage[english]{babel}
\usepackage{graphicx}
\usepackage{epsf}
\usepackage{pstricks}

\begin{document}

\title{Do L3 data indicate the existence of an isotensor meson ?
\footnote{Presented at Photon2005, Warsaw, by I.V.~Anikin.}}
\author{I.V. Anikin$^{a}$, B. Pire$^b$, O.V. Teryaev$^{a}$}
\affiliation{ $^a$Bogoliubov Laboratory of Theoretical Physics,
             JINR, 141980 Dubna, Russia\\
             $^b$CPHT, {\'E}cole Polytechnique,
             91128 Palaiseau Cedex, France\footnote{Unit{\'e} mixte 7644 du CNRS}}
\vspace{1.5cm}

\begin{abstract}
The QCD analysis of the hard exclusive production of $\rho^+\rho^-$ and
$\rho^0\rho^0$ mesons in two photon collisions shows that the recent experimental data
obtained by the L3 Collaboration at LEP can be understood as a signal
for the existence of an exotic isotensor resonance with a mass around $1.5\, {\rm GeV}$. 
\end{abstract}
\maketitle
  
\section{Introduction}
Exclusive reactions $\gamma^*\gamma\to A + B $ which may be accessed in
$e^+ e^-$ collisions have been shown \cite{DGPT} to have a partonic
interpretation in the kinematical region of large virtuality of one photon and
of small center of mass energy. The  scattering
amplitude factorizes in a long distance dominated object -- the
generalized distribution amplitude (GDA) -- and a short distance perturbatively
calculable scattering matrix. Data on the
$\rho^0 \rho^0$ and $\rho^+\rho^-$ channels have now been published \cite{L3}. 
Their analysis \cite{APT}  firstly shows
the compatibility of the QCD leading order analysis with experiment down to  quite modest
values of $Q^2$, and secondly enables to separate a twist-4 signal which may be interpreted as 
an isotensor contribution of potential great interest for the hadron spectroscopy.

\section{Framework}
Photon photon collisions may create a pair of isovector meson with total isospin $0$ or $2$.
Twist decomposition of meson production separates two quark operators from four quark operators.
The $\rho\rho$ state with $I=0$ can be projected on both the two and
four quark operators, while the state with $I=2$ on the four quark operator only.
Production of an exotic isotensor meson is thus a good probe of the validity of the
twist expansion. This may be contrasted with exotic hybrid meson production (with $J^{PC}
=1^{-+}$) which has a leading twist component \cite{hyb}. The flavour decomposition of correlators 
show this explicitly. For instance the  vacuum--to--$\rho\rho$ matrix element reads
\begin{eqnarray}
\label{cor2q}
\langle \rho^a \rho^b |
\bar\psi_f(0) \Gamma \psi_g(z) |0\rangle=
\delta^{ab} I_{fg}\Phi^{I=0}+
i\varepsilon^{abc}\tau^c_{fg}\Phi^{I=1},
\end{eqnarray}
where the quark fields are shown with 
flavour indices and $\Gamma$ stands for the corresponding $\gamma$-matrix.
The isoscalar and isovector GDA's $\Phi^{I}$ in (\ref{cor2q}) are unknown  (see
however \cite{BCDP} for their structure). 
The four quark GDA's $\tilde\Phi^{I,\,I_z=0}$ can be defined in an analogous way as
 the two quark GDA's.
Hence,  the amplitudes 
for $\rho^0\rho^0$ and $\rho^+\rho^-$ production in photon-photon collisions can be 
written under the form of the decomposition:
\begin{eqnarray}
\label{sumam}
{\cal A}_{(+,+)} = {\cal A}_{(+,+)\,2}^{I=0,\,I_z=0} +
{\cal A}_{(+,+)\,4}^{I=0,\,I_z=0} + {\cal A}_{(+,+)\,4}^{I=2,\,I_z=0},
\end{eqnarray}
where the subscripts $2$ and $4$  indicate that
the given amplitudes are associated with the two and four quark correlators, respectively.
The crucial point is that the amplitudes corresponding to $\rho^+\rho^-$ and $\rho^0\rho^0$ 
production are not independent :
\begin{eqnarray}
\label{rel1}
&&{\cal A}_{(+,+)\,k}^{I=0,\,I_z=0}(\gamma\gamma^*\to\rho^+\rho^-)=
{\cal A}_{(+,+)\,k}^{I=0,\,I_z=0}(\gamma\gamma^*\to\rho^0\rho^0)
\quad {\rm for}\,\,\, k=2,\,4
\nonumber\\
&&{\cal A}_{(+,+)\,4}^{I=2,\,I_z=0}(\gamma\gamma^*\to\rho^+\rho^-)=
-\frac{1}{2}{\cal A}_{(+,+)\,4}^{I=2,\,I_z=0}(\gamma\gamma^*\to\rho^0\rho^0).
\end{eqnarray}

\section{Cross sections and fitting procedure}
The production cross section   $\frac{d\sigma_{ee\to ee\rho^0\rho^0}}{dQ^2dW^2}$ may be written as :
\begin{eqnarray}
\label{xsec7}
&&
\frac{100\alpha^4}{9} G(S_{ee},Q^2,W^2) \beta
\Biggl(
f_{0}(W)\biggl[ {\bf S}^{I=0,I_3=0}_2+\frac{\alpha_S(Q^2)M^2_{R^0}}{Q^2}{\bf S}^{I=0,I_3=0}_4\biggr]^2
\Biggr.
\nonumber\\
\Biggl.
&& + f_{2}(W)\biggl[\frac{\alpha_S(Q^2)M^2_{R^2}}{Q^2}{\bf S}^{I=2,I_3=0}_4\biggr]^2+
\Biggr.
\\
\Biggl.
&&f_{02}(W)\biggl[ {\bf S}^{I=0,I_3=0}_2
+\frac{\alpha_S(Q^2)M^2_{R^0}}{Q^2}{\bf S}^{I=0,I_3=0}_4\biggr]
\frac{\alpha_S(Q^2)M^2_{R^2}}{Q^2}{\bf S}^{I=2,I_3=0}_4
\Biggr),
\nonumber
\end{eqnarray}
where $f_{i}(W)$ stand for the unknown but much interesting $W-$dependence of the GDAs
\cite{PS}, 
which we will parameterize with Breit-Wigner forms. The dimensionful structure constants
${\bf S}^{I,I_3=0}_4$ and ${\bf S}^{I=0,I_3=0}_2$ are related to the nonperturbative
vacuum--to--meson matrix elements and their relative magnitudes measures the importance 
of different twist components.
The function $G$ in (\ref{xsec7}) is the usual Weizsacker-Williams function.
The differential cross section corresponding to $\rho^+\rho^-$ production can be 
obtained using Eq. 3. The different weights of the twist-2 and twist 4-components then
enable to fit the  parameters ${\bf S}^{I,I_3=0}_4$ and ${\bf S}^{I=0,I_3=0}_2$ and the related 
functions $f_{i}(W)$.

We then fit the parameters associated with the
different twist contributions ; we get for the isoscalar sector
a  background described
as an "effective"  resonance with mass and width equal to 
$M_{R^0}=1.8\, {\rm GeV}$, $\Gamma_{R^0}=1.00\, {\rm GeV}$.
To describe data with  $0.2<Q^2<0.85$ GeV$^2$ (see Fig. 1)
we definitely need
an isotensor component which we parameterize with a Breit-Wigner representation of a 
resonance. The mass and widths of this resonance are then fitted as 
 $M_{R^2}=1.5\, {\rm GeV}$, $\Gamma_{R^2}=0.4\, {\rm GeV}$,  while
the parameters which measure the relative magnitudes of the amplitudes come out
as ${\bf S}^{I=0,I_3=0}_4 \in (0.002, \, 0.006)$ and ${\bf S}^{I=2,I_3=0}_4 \in (0.012, \, 0.018)$ 
, to be compared with 
${\bf S}^{I=0,I_3=0}_2 \in (0.12,\, 0.16)$. 

\begin{figure}[htb]
$$\includegraphics[width=7cm]{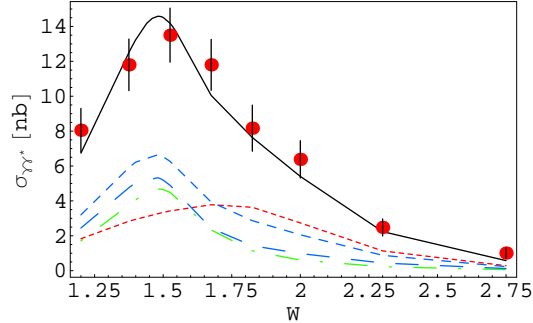}$$
\caption{$W$ dependence of the cross section $\sigma_{\gamma^*\gamma\to\rho^0\rho^0}$
 in the $0.2<Q^2<0.85$ region.
The short-dashed  (dash-dotted) line corresponds
to the twist $2$ (twist $4$) contribution, the middle-dashed and long-dashed lines to
 interference terms. Experimental data have been taken from \cite{L3}.}
\label{rho0rho0}
\end{figure}
As seen on Fig.2, the $Q^2-$dependence of both $\rho^0\rho^0$ and $\rho^+\rho^-$ production cross sections 
is then fairly well described on a wide range of values of $Q^2$. The leading twist amplitude is 
dominant almost down to  $Q^2= 1{\rm GeV}^2$ and the interference of twist 2 and twist 4
amplitudes is needed at lower values.

\begin{figure}[htb]
$$\includegraphics[width=7cm]{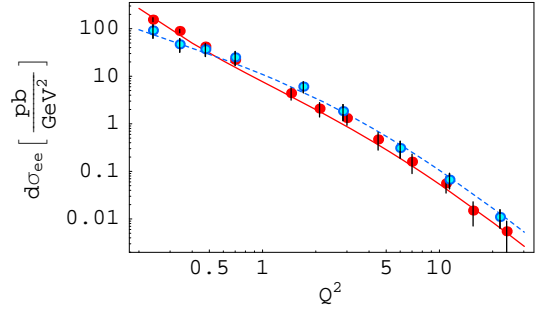}$$
\caption{The $Q^2$ dependence of the differential cross sections $d\sigma_{ee\to ee\rho^0\rho^0}/dQ^2$
and $d\sigma_{ee\to ee\rho^+\rho^-}/dQ^2$. The solid line corresponds
to the case of $\rho^0\rho^0$ production;
the dashed line to the case of $\rho^+\rho^-$ production. Experimental data have been taken from \cite{L3}.}
\label{rhorho1}
\end{figure}

\section{Conclusion}
The reaction $\gamma^* \gamma \to \rho \rho$ and its QCD analysis  thus proves its 
efficiency to reveal facts on hadronic physics which would remain quite difficult to explain in a quantitative way otherwise.  Other
aspects of QCD may be revealed in different kinematical regimes through the same reaction \cite{other}.
Its detailed experimental analysis at present intense electron colliders  and  in a future linear collider is thus extremely promising. 

We are grateful to N.N.~Achasov, A. Donnachie, J.~Field, M.~Kienzle, N.~Kivel, M.V.~Polyakov and I.~Vorobiev 
for useful discussions and correspondence. 
O.V.T. is indebted to Theory Division of CERN and CPHT, {\'E}cole Polytechnique, for warm hospitality. I.V.A.
expresses gratitude to Theory Division of CERN and University of Geneva for 
financial support of his visit.   
This work has been supported  in part by RFFI Grant 03-02-16816.

\end{document}